\documentclass[aps,preprint,prl]{revtex4-1}%
\usepackage{amssymb}
\usepackage{amsfonts}
\usepackage{amsmath}
\usepackage{graphicx}%
\providecommand{\U}[1]{\protect\rule{.1in}{.1in}}
\begin{document}
\title{ Nonequilibrium Spin Magnetization Quantum Transport Equations: Spin and
Charge Coupling}
\author{F. A. Buot }
\email{fbuot@gmu.edu, fbuot@clbri.org}
\affiliation{Computational Materials Science Center, MSN\ 6A2, George Mason University,
Fairfax, VA 22030, USA, }
\affiliation{C\&LB Research Institute (CLBRI), Carmen, Cebu, Philippines }
\author{R. A. Loberternos}
\author{R. E. S. Otadoy}
\author{D. L. Villarin}
\affiliation{Theoretical and Computational Sciences \& Engineering Group, Department of
Physics, University of San Carlos at Talamban, Cebu City, Philippines}

\begin{abstract}
The classical Bloch equations of spin magnetization transport is extended to
fully time-dependent and highly-nonlinear nonequilibrium spin magnetization
quantum distribution function transport (SMQDFT) equations. The relevant
variables are the spinor correlation functions which separate into charge and
spin magnetization distributions that becomes highly coupled in SMQDFT
equations. The leading terms consist of the Boltzmann kinetic equation with
spin-orbit coupling in a magnetic field together with spin-dependent
scattering terms which contribute to the torque. These do not have analogue
within the classical relaxation-dephasing picture, but are inherently quantum
many-body effects. These should incorporate the spatio--temporal-dependent
phase-space dynamics of Elliot-Yafet and D'yakonov-Perel scatterings. The
resulting SMQDFT equations should serve as a theoretical foundation for
computational spintronic and nanomagnetic device applications, in
ultrafast-switching-speed/low-power performance and reliability analyses.

\end{abstract}
\endpage{ }
\maketitle

\texttt{ }The ultrafast-switching-speed and power-dissipation performance
analyses of nanoelectronic devices has revealed the inadequacy of the
classical Boltzmann transport equation (BTE) and ushered its extension of the
to fully time-dependent and highly-nonlinear nonequilibrium quantum
distribution function (QDF) transport equations for charge carriers. This has
been achieved through the use of non-equilibrium Green's function, obtained
either by the time-contour quantum field formulation of
Schwinger\cite{schwinger}, Keldysh\cite{keldysh}, and Kadanoff and
Baym\cite{kadanoff}, or by the real-time quantum superfield formulation of
Buot coupled with his lattice Weyl (LW) transformation
technique\cite{lwww,LaRNC, buot1, buot2}. The extension of BTE to QDF
transport equation has proved to be highly crucial in discovering autonomous
current oscillations in resonant tunneling devices through numerical
simulations, and in resolving controversial issues concerning highly-nonlinear
and bistable current-voltage characteristics found in the
experiments\cite{buotjensen, jbpaper}. Indeed, the QDF approach in phase space
has so far been the most successful technique in the time-dependent analyses
of open and active nanosystem and nanodevices\cite{jbpaper}.

To the authors' knowledge, the extension of the well-known classical Bloch
equation for spin transport, which is the analogue of the classical BTE for
particle transport, to fully time-dependent and highly-nonlinear
nonequilibrium QDF transport equations for the magnetization has not been
reported. With the exploding surge of interest on spintronics and
nanomagnetics\cite{wolf}, there is an urgent need for this fully quantum
transport extension of the classical Bloch equation to guide the numerical
simulation of the speed-power switching performance of realistic spin
nanostructures/transistors. The well-known classical Bloch equations for spin
systems is usually given in the diffusive regime\cite{bediff} as%
\begin{align}
\frac{\partial M_{x}}{\partial t}  &  =\left[  D\nabla^{2}M_{x}-\frac{M_{x}%
}{T_{2}}\right]  +\gamma\left(  \vec{M}\times\vec{B}\right)  _{x},\nonumber\\
\frac{\partial M_{y}}{\partial t}  &  =\left[  D\nabla^{2}M_{y}-\frac{M_{y}%
}{T_{2}}\right]  +\gamma\left(  \vec{M}\times\vec{B}\right)  _{y},\nonumber\\
\frac{\partial M_{z}}{\partial t}  &  =\left[  D\nabla^{2}M_{z}-\frac
{M_{z}-M_{o}}{T_{1}}\right]  +\gamma\left(  \vec{M}\times\vec{B}\right)  _{z},
\label{classBlochEq}%
\end{align}
where $\gamma$ is the gyromagnetic ratio, $T_{1}$ is the spin-relaxation
characteristic time, and $T_{2}$ is the spin-dephasing characteristic time.

The purpose of this paper is to extend this classical Bloch equations to a
fully time-dependent and highly-nonlinear nonequilibrium SMQDFT equations on
the same level as in the extension of the classical BTE to full QDF transport
equation based on nonequilibrium Green's function technique\cite{schwinger,
keldysh, kadanoff, LaRNC, buot1, buot2}. In this paper we will first treat
spin transport within a single energy band. The multiband spin transport,
where pseudo-spin plays a role, will be reported in a separate arxiv.org
online paper.

Our starting point is the general quantum transport expressions for fermions,
as obtained from the real-time quantum superfield theoretical
technique\cite{buot1,buot2}. Upon dropping the Cooper pairing terms in those
expressions which do not concern us in this paper (their corresponding
transport equations\cite{buot1,buot2} are important in nonequilibrium
superconductivity), the nonequilibrium Green's function transport equations
reduce to the following known expressions\cite{schwinger, keldysh, kadanoff},
\begin{align}
i\hbar\left(  \frac{\partial}{\partial t_{1}}+\frac{\partial}{\partial t_{2}%
}\right)  G^{\gtrless}  &  =\left[  vG^{\gtrless}-G^{\gtrless}v^{T}\right]
\nonumber\\
&  +\left[  \Sigma^{r}G^{\gtrless}-G^{\gtrless}\Sigma^{a}\right]  +\left[
\Sigma^{\gtrless}G^{a}-G^{r}\Sigma^{\gtrless}\right]  . \label{grnlesseq}%
\end{align}

The quantum distribution-function transport equations in $\left(  p,q\right)
=\left(  \mathbf{p},\mathbf{q},E,t\right)  $-space are obtained by applying
the LW transformation\cite{lwww} (although continuum approximation is employed
in this paper, this is not essential and we use the word \textquotedblright
lattice\textquotedblright\ when referring to solid-state problems)\cite{buot1,
buot2}. The LW transform $a(p,q)$ of any operator $\hat{A}$\ is defined by the
following relation
\begin{equation}
a(p,q)=\int dv\exp\left(  \frac{ip\cdot v}{\hbar}\right)  \left\langle
q-\frac{v}{2}\right\vert \hat{A}\left\vert q+\frac{v}{2}\right\rangle
\equiv\mathcal{W\ }A\left(  \mathbf{v},\tau,\mathbf{q},t\right)  ,
\label{preq142}%
\end{equation}
where the matrix element in the integrand is evaluated between two Wannier
functions, and $\mathcal{W}$ indicates the operation of taking the LW
transform. This transformation enables the numerical simulation of real
time-dependent dynamical open systems by transforming from \textit{two-point}
correlation functions to numerically manageable and physically meaningful
\textit{local}\ functions in phase space\cite{contrast}. The well-known
electron Wigner distribution function (WDF), $f_{w}\left(  \mathbf{p}%
,\mathbf{q},t\right)  $, is obtained from the correlation function $-i\hbar
G^{<}$ by performing integration over the energy variable
\begin{equation}
f_{w}\left(  \mathbf{p},\mathbf{q},t\right)  =\frac{1}{2\pi\hbar}\int
dE\ \left(  -i\hbar G^{<}\left(  \mathbf{p},E,\mathbf{q},t\right)  \right)  .
\label{preq147}%
\end{equation}

In the absence of particle pairing of superconductivity to obtain Eq.
(\ref{grnlesseq}), we may also write the general transport equation,written
for '$<$'-quantities as [here $\left\{  A,B\right\}  $ and $\left[
A,B\right]  $ means anticommutator and commutator, respectively, of $A$ and
$B$],%
\begin{align}
&  i\hbar\left(  \frac{\partial}{\partial t_{1}}+\frac{\partial}{\partial
t_{2}}\right)  G^{<}\nonumber\\
&  =\left[  \mathcal{\tilde{H}},G^{<}\right]  +\left[  \Sigma^{<}%
,\operatorname{Re}G^{r}\right]  -\frac{i}{2}\left\{  \Gamma,G^{<}\right\}
+\frac{i}{2}\left\{  \Sigma^{<},A\right\}  , \label{QTE}%
\end{align}
where the single-particle Hamiltonian,
\[
\mathcal{\tilde{H}=H}_{o}+\Sigma^{\delta}+\operatorname{Re}\Sigma^{r},
\]
with $\Sigma^{\delta}$ being the 'singular part' (delta function in time) of
the self-energy.
\[
\mathcal{H}_{o}=E\left(  p\right)  +V\left(  q\right)  ,
\]
where $E\left(  p\right)  $ is the electron dispersion relation or energy-band
function and $V\left(  q\right)  $ is the self-consistent single-particle
potential function. In obtaining Eq. (\ref{QTE}) from Eq. (\ref{grnlesseq}),
we made use of the relations, $\left(  G^{a}\right)  ^{\dagger}=G^{r}$,
and\ \ $\Sigma^{<\dagger}\left(  1,2\right)  =-\Sigma^{<}\left(  2,1\right)
$, and the following relations, $iA\left(  1,2\right)  =-2i\operatorname{Im}%
G^{r}=-\left(  G^{>}\left(  1,2\right)  -G^{<}\left(  1,2\right)  \right)  $,
$i\Gamma\left(  1,2\right)  =-2i\operatorname{Im}\Sigma^{r}=-\left(
\Sigma^{>}\left(  1,2\right)  -\Sigma^{<}\left(  1,2\right)  \right)  $.

The second term in Eq. (\ref{QTE}) is a commutator and also describes
evolution in phase-space although only serving to complicate, by virtue of the
scattering term,$\ \Sigma^{<}$, the kinetics of particle motion described by
the first term, a sort of complex interference phenomena similar to
\textit{zitterbewegung} in Dirac quantum mechanics\cite{bellstheorem}.
Although often neglected in considering quantum transport of particles as this
is expected to make a small corrections to the particle current, we will
include the $\left[  \Sigma^{<},\operatorname{Re}G^{r}\right]  $ term in our
present considerations for completeness. Note that in the gradient expansion
of Eq. (\ref{QTE}), with the $\left[  \Sigma^{<},\operatorname{Re}%
G^{r}\right]  $ term neglected, the leading terms immediately give the
classical Boltzmann transport equation for spinless systems\cite{boltzmann}.
However, even this is no longer quite true for spin system as we shall see
later, since the spinor scattering terms contribute to the torque in the system.

The third term of Eq. (\ref{QTE}), which is proportional to $G^{<}$, is the
\textit{scattering-out} term, and the last term which is proportional to the
spectral density is the \textit{scattering-in} term, in analogy with the
collision terms in the classical Boltzmann equation. The collision terms can
also be rewritten using the identity,
\[
-\frac{i}{2}\left\{  \Gamma,G^{<}\right\}  +\frac{i}{2}\left\{  \Sigma
^{<},A\right\}  =-\frac{1}{2}\left\{  \Sigma^{<},G^{>}\right\}  +\frac{1}%
{2}\left\{  \Sigma^{>},G^{<}\right\}  ,
\]
which also leads to $i\Sigma^{>}$ and $-i\Sigma^{<}$ terms, upon performing
the $\mathcal{W}$ operation, as scattering-out and scattering-in collision
terms, respectively\cite{kadanoff}.

The equation for $G^{r}$ and $G^{a}=G^{r\dagger}$ are simpler and are given
as, $\hat{G}^{-1}G^{r}=0,$ $G^{a}\left(  \hat{G}^{-1}\right)  ^{\dagger}=0,$
where
\[
\hat{G}^{-1}G^{r}\left(  1,2\right)  =i\hbar\left(  \frac{\partial}{\partial
t_{1}}+\frac{\partial}{\partial t_{2}}\right)  G^{r}\left(  1,2\right)
-\left[  \mathcal{\tilde{H}}+\Sigma^{r},G^{r}\right]  \left(  1,2\right)  .
\]

The above equations are then cast in the language of phase-space QDF transport
equations, similar to that of the classical Boltzmann equation by the use of
\ Buot lattice Weyl-Wigner formulation of the band dynamics of electrons in a
solid\cite{lwww}. In the \textit{continuum approximation}, the first thing to
do is to recast the \textit{two-point} space and time arguments in all the
transport equations as follows: $t_{1}=t-\frac{\tau}{2},$ $t_{2}=t+\frac{\tau
}{2},$ $\mathbf{q}_{1}=\mathbf{q}-\frac{\mathbf{v}}{2},$ $\mathbf{q}%
_{2}=\mathbf{q}+\frac{\mathbf{v}}{2}.$ We also need to define a ($3+1$%
)-dimensional canonical variables, namely, $p=\left(  \mathbf{p},-E\right)  ,$
$q=\left(  \mathbf{q},t\right)  ,$ which allows us to write $v=\left(
\mathbf{v},\tau\right)  $. \ The transformation of Eq. (\ref{preq142}) is then applied.

For simplicity in spin quantum transport equations, we consider only a single
energy band having spin indices, $\downarrow$ and $\uparrow$. Incorporating
the $\downarrow$ and $\uparrow$ variables in Eqs. (\ref{grnlesseq}) and
(\ref{QTE}), we obtain exactly four transport equations for $G_{\uparrow
\uparrow}^{<}\left(  1,2\right)  $, $G_{\uparrow\downarrow}^{<}\left(
1,2\right)  $, $G_{\downarrow\uparrow}^{<}\left(  1,2\right)  $, and
$G_{\downarrow\downarrow}^{<}\left(  1,2\right)  $. In spintronics, we are
interested in the time-dependent evolution of the magnetic polarization,
$S_{z}=i\hbar\left(  G_{\uparrow\uparrow}^{<}-G_{\downarrow\downarrow}%
^{<}\right)  ,$ as this is transported across the device. In order to achieve
this, we need to perform \textit{linearly-independent} combinations of the
components of the spinor nonequilibrium Green's function consisting of the
set: $\left\{  G_{\uparrow\uparrow}^{<}\left(  1,2\right)  ,G_{\uparrow
\downarrow}^{<}\left(  1,2\right)  ,G_{\downarrow\uparrow}^{<}\left(
1,2\right)  ,G_{\downarrow\downarrow}^{<}\left(  1,2\right)  \right\}  $. The
resulting new set will become the relevant and independent correlation
functions more pertinent to the spin-transport problem, which consists of the
separation into charge distribution and spin magnetization vector.

The separation in terms of a scalar and vector representing the total charge
and spin-vector correlation functions, naturally occurs as expansion
coefficient in terms of the Pauli matrices,%
\begin{equation}
\left(
\begin{array}
[c]{cc}%
G_{\uparrow\uparrow}^{<} & G_{\uparrow\downarrow}^{<}\\
G_{\downarrow\uparrow}^{<} & G_{\downarrow\downarrow}^{<}%
\end{array}
\right)  =\frac{1}{2}\left(  S_{o}\hat{I}+\vec{S}\cdot\vec{\sigma}\right)  ,
\label{expandG_sigmas}%
\end{equation}
where $\hat{I}$ is the $2\times2$ identity matrix, and%
\begin{equation}
\left.
\begin{array}
[c]{c}%
S_{x}=\left(  G_{\downarrow\uparrow}^{<}+G_{\uparrow\downarrow}^{<}\right)
,\\
iS_{y}=\left(  G_{\downarrow\uparrow}^{<}-G_{\uparrow\downarrow}^{<}\right)
,\\
S_{z}=\left(  G_{\uparrow\uparrow}^{<}-G_{\downarrow\downarrow}^{<}\right)
,\\
S_{o}=\left(  G_{\uparrow\uparrow}^{<}+G_{\downarrow\downarrow}^{<}\right)  ,
\end{array}
\right\}  , \label{Sdefine1}%
\end{equation}
where we drop the '$<$' superscript in the spin correlation functions, $S_{j}%
$. In the frame of reference where the $z$ direction is fixed by the magnetic
field, the $S_{x}$ and $S_{y}$ evolution equation describe \textit{dephasing}
mechanisms in the $x$-$y$ plane of the 'Bloch sphere'. The total charge of the
system, $\rho\left(  p,q,t\right)  $ is represented by the scalar $S_{o}$
i.e., from Eq. (\ref{preq147}) we have,
\begin{equation}
\rho\left(  \vec{q},t\right)  =\frac{1}{\left(  2\pi\hbar\right)  ^{2}}%
%TCIMACRO{\dint }%
%BeginExpansion
{\displaystyle\int}
%EndExpansion
d\vec{p}\ dE\left(  -i\hbar S_{o}\left(  \vec{p},E,\vec{q},t\right)  \right)
, \label{Eqp}%
\end{equation}
where $S_{o}\left(  \vec{p},E,\vec{q},t\right)  $ is the LW of correlation
$S_{o}$. Thus Eq. (\ref{expandG_sigmas}) performs the separation of the charge
and spin correlation functions as the relevant variables of any fermion system
with spin degree of freedom.

For slowly varying system where the total charge vary slowly with space and
time, perhaps in some spin transfer torque systems, we can reduce the number
of important spin correlation functions in the spin quantum transport problem
by making an approximation to $S_{o}$\cite{atomic}. Assuming the conservation
of particles possessing spin degree of freedom within semiconductor channel
of, say, a spintronic transfer torque transistor, we must have the following
relation for the '$\uparrow\uparrow$' and '$\downarrow\downarrow$' rates of
change, $\frac{\partial\rho_{\uparrow\uparrow}^{<}\left(  p,q,t\right)
}{\partial t}=-\frac{\partial\rho_{\downarrow\downarrow}^{<}\left(
p,q,t\right)  }{\partial t}$, where $\rho_{\uparrow\uparrow}\left(
p,q,t\right)  $ and $\rho_{\downarrow\downarrow}\left(  p,q,t\right)  $ are
the LW transforms of $i\hbar G_{\uparrow\uparrow}^{<}$ and $i\hbar
G_{\downarrow\downarrow}^{<}$, respectively.

Generally however, $S_{o}$ will be varying in both space and time, i.e., a
two-point nonequilibrium correlation function in space and time, describing
the varying charge within the spintronic device due to depletion and
accumulation in various regions within the device, such as in spin transfer
torque devices employing thin barriers between the channel and ferromagnetic leads.

We can also write the spinor self-energies used in Eq. (\ref{grnlesseq})
as\cite{note1, note2} into spin-independent scalar [denoted by overscript bar
symbol] and vector quantities,
\[
\Sigma_{\alpha\beta}^{a,r,<}=\frac{1}{2}\left(  \bar{\Sigma}^{a,r,<}\hat
{I}+\vec{\Xi}^{a,r,<}\cdot\vec{\sigma}\right)  ,
\]
which we write explicitly as,%
\[
\Sigma_{\alpha\beta}^{r}=\frac{1}{2}\left(
\begin{array}
[c]{cc}%
\bar{\Sigma}^{r}+\Xi_{z}^{r} & \Xi_{x}^{r}-i\Xi_{y}^{r}\\
\Xi_{x}^{r}+i\Xi_{y}^{r} & \bar{\Sigma}^{r}-\Xi_{z}^{r}%
\end{array}
\right)  ,
\]%
\[
\Sigma_{\alpha\beta}^{a}=\frac{1}{2}\left(
\begin{array}
[c]{cc}%
\bar{\Sigma}^{a}+\Xi_{z}^{a} & \Xi_{x}^{a}-i\Xi_{y}^{a}\\
\Xi_{x}^{a}+i\Xi_{y}^{a} & \bar{\Sigma}^{a}-\Xi_{z}^{a}%
\end{array}
\right)  ,
\]%
\[
\Sigma_{\alpha\beta}^{<}=\frac{1}{2}\left(
\begin{array}
[c]{cc}%
\bar{\Sigma}^{<}+\Xi_{z}^{<} & \Xi_{x}^{<}-i\Xi_{y}^{<}\\
\Xi_{x}^{<}+i\Xi_{y}^{<} & \bar{\Sigma}^{<}-\Xi_{z}^{<}%
\end{array}
\right)  .
\]

Likewise, the spinor quantities used in Eq. (\ref{QTE}), namely, the
spin-dependent \textit{scattering-out }$\Gamma$term is cast in the form,
\begin{align}
\frac{\Gamma_{\alpha\beta}}{2}  &  =\frac{1}{2}\left(  \bar{\Gamma}\hat
{I}+\vec{\gamma}\cdot\vec{\sigma}\right)  =-\operatorname{Im}\Sigma
_{\alpha\beta}^{r}\nonumber\\
&  =-\frac{1}{2}\left(  \operatorname{Im}\bar{\Sigma}^{r}\hat{I}%
+\operatorname{Im}\vec{\Xi}^{r}\cdot\vec{\sigma}\right)  , \label{GammaSep}%
\end{align}
so that%
\begin{align}
\bar{\Gamma}  &  =-\operatorname{Im}\bar{\Sigma}^{r},\nonumber\\
\vec{\gamma}  &  =-\operatorname{Im}\vec{\Xi}^{r}. \label{GammaSep2}%
\end{align}
Similarly, we have for the spin-dependent spectral function, $A$, in the
\textit{scattering-in} term as
\begin{align}
\frac{A_{\alpha\beta}}{2}  &  =\frac{1}{2}\left(  \bar{A}\hat{I}%
+\mathcal{\vec{A}}\cdot\vec{\sigma}\right)  =-\operatorname{Im}S_{\alpha\beta
}^{r}\nonumber\\
&  =-\frac{1}{2}\left(  \operatorname{Im}\bar{S}^{r}\hat{I}+\operatorname{Im}%
\vec{S}^{r}\cdot\vec{\sigma}\right)  , \label{SpectralSep}%
\end{align}
so that%
\begin{align}
\bar{A}  &  =-\operatorname{Im}\bar{S}^{r},\nonumber\\
\mathcal{\vec{A}}  &  \mathcal{=-}\operatorname{Im}\vec{S}^{r}.
\label{SpectralSep2}%
\end{align}

In the effective-mass approximation, we may take the corresponding
single-particle Hamiltonian $\mathcal{\tilde{H}}_{s}$ for spin systems in a
magnetic field with spin-orbit coupling as
\begin{equation}
\mathcal{\tilde{H}}_{s}\mathcal{=}\left(  \frac{\mathcal{\hat{P}}^{2}%
}{2m^{\ast}}-eV\left(  q\right)  +\frac{1}{2}\operatorname{Re}\bar{\Sigma}%
^{r}\right)  \hat{I}+H_{so}\left(  \mathcal{\hat{P}},\hat{Q}\right)  +\frac
{1}{2}\left(  \operatorname{Re}\vec{\Sigma}^{r}\cdot\vec{\sigma}-g_{eff}%
\mu_{B}\vec{B}\cdot\vec{\sigma}\right)  , \label{hamiltonian}%
\end{equation}
where $\mathcal{\hat{P}}=\hat{P}+\frac{e}{c}A\left(  q\right)  $, electric
charge, $e=\left\vert e\right\vert $, $V\left(  q\right)  $ is the
self-consistent applied potential, and $\vec{B}=\nabla_{\vec{q}}\times
A\left(  q\right)  $. The second term, $H_{so}\left(  \mathcal{\hat{P}}%
,\hat{Q}\right)  ,$ is the spin-orbit coupling which may consists of
Dresselhaus and/or Rashba term\cite{dresRashba} as well as transistor-gate and
strain-induced spin-orbit coupling. The term $\operatorname{Re}\vec{\Sigma
}^{r}\cdot\vec{\sigma}$ comes from the spin vector of the spin-dependent part
of $\operatorname{Re}\Sigma_{\alpha\beta}^{r}$ in the single particle
Hamiltonian. Its presence basically modifies the effective magnetic field due
to external fields and spin-orbit coupling.

We have in general,
\[
H_{so}\left(  \mathcal{\hat{P}},\hat{Q}\right)  =\mu_{B}\frac{\vec{q}%
\times\mathcal{\hat{P}}}{2\hbar m^{\ast}c^{2}}\left\vert \frac{\vec{E}}%
{q}\right\vert \cdot\vec{\sigma},
\]
which may acquire definite form in lower dimensional system based on symmetry
considerations\cite{dresRashba}. In what follows, we write the most general
form of the single-particle Hamiltonian, $\mathcal{\tilde{H}}_{s}$, as
\[
\mathcal{\tilde{H}}_{s}=\frac{1}{2}\left(  \bar{H}\left(  \hat{P},\hat
{Q}\right)  \hat{I}-\vec{h}\left(  \hat{P},\hat{Q}\right)  \cdot\vec{\sigma
}\right)  ,
\]
where $\vec{h}\left(  \hat{P},\hat{Q}\right)  =\left(  g_{s}\frac{e}{2mc}%
\vec{B}_{eff}-\operatorname{Re}\vec{\Sigma}^{r}\right)  $ . Therefore
\begin{equation}
\mathcal{\tilde{H}}_{s}=\frac{1}{2}\left(  \bar{H}\ \hat{I}-\vec{\sigma}%
\cdot\mathcal{\vec{B}}\right)  \text{,} \label{spinH2}%
\end{equation}
where
\begin{equation}
\mathcal{\vec{B}=}g_{s}\frac{e}{2mc}\vec{B}_{eff}, \label{spinB}%
\end{equation}
and for electron-spin angular momentum, the \textit{g-factor}, $g_{s}\simeq2$.
Here the effective $\vec{B}_{eff}\left(  \hat{P},\hat{Q}\right)  $ includes
the effects of transistor-gate or strain induced spin-orbit coupling, and
applied magnetic fields. $H\left(  \hat{P},\hat{Q}\right)  $ consists of the
energy-band structure and spin-independent confining potential due to the
device heterostructure coupled with the applied bias at the terminals. The
spin-orbit coupling in $\vec{B}_{eff}\left(  \hat{P},\hat{Q}\right)  $ also
depends very much on the electric field of the self-consistent potential
inside the device. This in turn depends on the total electric charge,
represented by $S_{o}$, through the Poisson equation. Although we could
include $\operatorname{Re}\vec{\Sigma}^{r}$ in the definition of
$\mathcal{\vec{B}}$ above, we treat this separately to emphasize the possible
role of scattering terms in contributing to the torque in the system.

The resulting equations obeyed by the spin correlation functions defined by
Eq. (\ref{Sdefine1}) in accordance with Eq. (\ref{grnlesseq}) can be written
in vector notation as,%
\begin{align}
&  i\hbar\left(  \frac{\partial}{\partial t_{1}}+\frac{\partial}{\partial
t_{2}}\right)  \vec{S}^{<}\nonumber\\
&  =\frac{1}{2}\left[  \bar{H},\vec{S}^{<}\right]  +\frac{1}{2}\left[
\bar{\Sigma}^{r}\vec{S}^{<}-\ \vec{S}^{<}\bar{\Sigma}^{a}\right]  +\frac{1}%
{2}\left[  \bar{\Sigma}^{<}\vec{S}^{a}-\ \vec{S}^{r}\bar{\Sigma}^{<}\right]
\nonumber\\
&  +\frac{i}{2}\left[  \mathcal{\vec{B}}\times\vec{S}^{<}-\vec{S}^{<}%
\times\mathcal{\vec{B}}\right]  \nonumber\\
&  +i\frac{1}{2}\left[  \vec{\Xi}^{r}\ \times\vec{S}^{<}-\ \vec{S}^{<}%
\times\vec{\Xi}^{a}\right]  +i\frac{1}{2}\left[  \vec{\Xi}^{<}\times\vec
{S}^{a}-\ \vec{S}^{r}\times\vec{\Xi}^{<}\right]  \nonumber\\
&  +\frac{1}{2}\left[  \mathcal{\vec{B}},S_{o}^{<}\right]  +\frac{1}{2}\left[
\vec{\Xi}^{r}S_{o}^{<}-S_{o}^{<}\vec{\Xi}^{a}\right]  +\frac{1}{2}\left[
\vec{\Xi}^{<}S_{o}^{a}-S_{o}^{r}\vec{\Xi}^{<}\right]  ,\label{singleBand}%
\end{align}
and for the \textit{scalar} correlation function representing the total charge
,
\begin{align}
&  i\hbar\left(  \frac{\partial}{\partial t_{1}}+\frac{\partial}{\partial
t_{2}}\right)  S_{o}^{<}\nonumber\\
&  =\frac{1}{2}\left[  \bar{H},S_{o}^{<}\right]  +\frac{1}{2}\left[
\bar{\Sigma}^{r}S_{o}^{<}-\ S_{o}^{<}\bar{\Sigma}^{a}\right]  +\frac{1}%
{2}\left[  \bar{\Sigma}^{<}S_{o}^{a}-\ S_{o}^{r}\bar{\Sigma}^{<}\right]
\nonumber\\
&  +\frac{1}{2}\left[  \mathcal{\vec{B}}\cdot\vec{S}^{<}-\vec{S}^{<}%
\cdot\mathcal{\vec{B}}\right]  +\frac{1}{2}\left[  \vec{\Xi}^{r}\cdot\vec
{S}^{<}-\vec{S}^{<}\cdot\vec{\Xi}^{a}\right]  \ \ \nonumber\\
&  +\frac{1}{2}\left[  \vec{\Xi}^{<}\cdot\vec{S}^{a}-\vec{S}^{r}\cdot\vec{\Xi
}^{<}\right]  .\ \ \label{scalarEq}%
\end{align}
For self-consistency, one must also be solved for potential, $\Phi$, or
electric field, $\vec{E},$ using the Poisson equation,%
\[
\nabla^{2}\Phi=e\frac{\rho\left(  \vec{q},t\right)  -\rho_{o}}{\varepsilon
_{o}},
\]
where $\Phi=V$ in Eq. (\ref{hamiltonian}), $\rho_{o}$ is the positive
background charge and $\rho\left(  \vec{q},t\right)  $ is given by Eq.
(\ref{Eqp}).

We also have the equivalent equation for $\vec{S}^{<}$,
\begin{align}
&  i\hbar\left(  \frac{\partial}{\partial t_{1}}+\frac{\partial}{\partial
t_{2}}\right)  \vec{S}^{<}\nonumber\\
&  =\frac{1}{2}\left[  \bar{H}+\operatorname{Re}\bar{\Sigma}^{r},\vec{S}%
^{<}\right]  +\frac{i}{2}\left\{  \operatorname{Im}\bar{\Sigma}^{r},\vec
{S}^{<}\right\}  -\frac{i}{2}\left\{  \bar{\Sigma}^{<},\operatorname{Im}%
\vec{S}^{r}\right\}  +\left[  \bar{\Sigma}^{<},\operatorname{Re}\vec{S}%
^{r}\right] \nonumber\\
&  +\frac{i}{2}\left[  \left(  \mathcal{\vec{B}}+\operatorname{Re}\vec{\Xi
}^{r}\right)  \times\vec{S}^{<}-\vec{S}^{<}\times\left(  \mathcal{\vec{B}%
}+\operatorname{Re}\vec{\Xi}^{r}\right)  \right] \nonumber\\
&  -\frac{1}{2}\left\{  \operatorname{Im}\vec{\Xi}^{r}\ \times\vec{S}%
^{<}+\ \vec{S}^{<}\times\operatorname{Im}\vec{\Xi}^{r}\right\}  \ +\frac{1}%
{2}\left\{  \vec{\Xi}^{<}\times\operatorname{Im}\vec{S}^{r}+\operatorname{Im}%
\ \vec{S}^{r}\times\vec{\Xi}^{<}\right\} \nonumber\\
&  +\frac{i}{2}\left[  \vec{\Xi}^{<}\times\operatorname{Re}\vec{S}%
^{r}-\ \operatorname{Re}\vec{S}^{r}\times\vec{\Xi}^{<}\right] \nonumber\\
&  +\frac{1}{2}\left[  \mathcal{\vec{B}},S_{o}^{<}\right]  +\frac{1}{2}\left[
\operatorname{Re}\vec{\Xi}^{r},S_{o}^{<}\right]  +\frac{i}{2}\left\{
\operatorname{Im}\vec{\Xi}^{r},S_{o}^{<}\right\}  +\left[  \vec{\Xi}%
^{<},\operatorname{Re}S_{o}^{r}\right]  -\frac{i}{2}\left\{  \vec{\Xi}%
^{<},\operatorname{Im}S_{o}^{r}\right\}  , \label{equivEq}%
\end{align}
which can be reduced to that which one can derive from Eq. (\ref{QTE}), by
using the relations in Eqs. (\ref{GammaSep2}) and (\ref{SpectralSep2}). The
result is%

\begin{align}
&  i\hbar\left(  \frac{\partial}{\partial t_{1}}+\frac{\partial}{\partial
t_{2}}\right)  \vec{S}^{<}\nonumber\\
&  =\frac{1}{2}\left[  \bar{H}+\operatorname{Re}\bar{\Sigma}^{r},\vec{S}%
^{<}\right]  -\frac{i}{2}\left\{  \bar{\Gamma},\vec{S}^{<}\right\}  +\frac
{i}{2}\left\{  \bar{\Sigma}^{<},\mathcal{\vec{A}}\right\}  +\left[
\bar{\Sigma}^{<},\operatorname{Re}\vec{S}^{r}\right] \nonumber\\
&  +\frac{i}{2}\left[  \left(  \mathcal{\vec{B}}+\operatorname{Re}\vec{\Xi
}^{r}\right)  \times\vec{S}^{<}-\vec{S}^{<}\times\left(  \mathcal{\vec{B}%
}+\operatorname{Re}\vec{\Xi}^{r}\right)  \right] \nonumber\\
&  +\frac{1}{2}\left\{  \vec{\gamma}\ \times\vec{S}^{<}+\ \vec{S}^{<}%
\times\vec{\gamma}\right\}  \ -\frac{1}{2}\left\{  \vec{\Xi}^{<}%
\times\mathcal{\vec{A}}+\mathcal{\vec{A}}\times\vec{\Xi}^{<}\right\}
\nonumber\\
&  +\frac{i}{2}\left[  \vec{\Xi}^{<}\times\operatorname{Re}\vec{S}%
^{r}-\ \operatorname{Re}\vec{S}^{r}\times\vec{\Xi}^{<}\right] \nonumber\\
&  +\frac{1}{2}\left[  \left(  \mathcal{\vec{B}+}\operatorname{Re}\vec{\Xi
}^{r}\right)  ,S_{o}^{<}\right]  +\left[  \vec{\Xi}^{<},\operatorname{Re}%
S_{o}^{r}\right]  -\frac{i}{2}\left\{  \vec{\gamma},S_{o}^{<}\right\}
-\frac{i}{2}\left\{  \vec{\Xi}^{<},\operatorname{Im}S_{o}^{r}\right\}  .
\label{equivEq2}%
\end{align}
Note that our definition of $\mathcal{\vec{B}}$, Eq. (\ref{spinB}), has the
unit of energy, just like the self-energies.

The equivalent expression derived from Eq. (\ref{QTE}) in terms of the
magnetization vector, $\mathcal{\vec{M}}=g_{s}\frac{e\hbar}{2m^{\ast}%
c}\mathcal{\vec{S}=}g_{s}\mu_{B}$ (the $g$-factor for spin, $g_{s}\simeq2$)
is,%
\begin{align}
&  i\hbar\left(  \frac{\partial}{\partial t_{1}}+\frac{\partial}{\partial
t_{2}}\right)  \mathcal{\vec{M}}\nonumber\\
&  =\frac{1}{2}\left[  \bar{H}+\operatorname{Re}\bar{\Sigma}^{r}%
,\mathcal{\vec{M}}\right]  -\frac{i}{2}\left\{  \bar{\Gamma},\mathcal{\vec{M}%
}\right\}  +\frac{i}{2}g_{s}\mu_{B}\left\{  \bar{\Sigma}^{<},\mathcal{\vec{A}%
}\right\}  +\frac{1}{2}\left[  \bar{\Sigma}^{<},\operatorname{Re}%
\mathcal{\vec{M}}^{r}\right] \nonumber\\
&  +\frac{i}{2}\left[  \left(  \mathcal{\vec{B}}+\operatorname{Re}\vec{\Xi
}^{r}\right)  \times\mathcal{\vec{M}}-\mathcal{\vec{M}}\times\left(
\mathcal{\vec{B}}+\operatorname{Re}\vec{\Xi}^{r}\right)  \right] \nonumber\\
&  +\frac{1}{2}\left\{  \vec{\gamma}\ \times\mathcal{\vec{M}}+\ \mathcal{\vec
{M}}\times\vec{\gamma}\right\}  \ -\frac{1}{2}g_{s}\mu_{B}\left\{  \vec{\Xi
}^{<}\times\mathcal{\vec{A}}+\mathcal{\vec{A}}\times\vec{\Xi}^{<}\right\}
\nonumber\\
&  +\frac{i}{2}\left[  \vec{\Xi}^{<}\times\operatorname{Re}\mathcal{\vec{M}%
}^{r}-\ \operatorname{Re}\mathcal{\vec{M}}^{r}\times\vec{\Xi}^{<}\right]
\nonumber\\
&  +\frac{1}{2}\left[  \left(  \mathcal{\vec{B}+}\operatorname{Re}\vec{\Xi
}^{r}\right)  ,M_{o}^{<}\right]  +\frac{1}{2}\left[  \vec{\Xi}^{<}%
,\operatorname{Re}M_{o}^{r}\right]  -\frac{i}{2}\left\{  \vec{\gamma}%
,M_{o}^{<}\right\}  -\frac{i}{2}\left\{  \vec{\Xi}^{<},\operatorname{Im}%
M_{o}^{r}\right\}  , \label{magTransportEq2}%
\end{align}
where we defined,%
\begin{align*}
\mathcal{\vec{M}}^{r}  &  =g_{s}\mu_{B}\vec{S}^{r},\\
M_{o}^{<}  &  =g_{s}\mu_{B}S_{o}^{<}.
\end{align*}
First, let us evaluate Eqs. (\ref{magTransportEq2}) by assuming constant
magnetic field $\mathcal{B}$ and in the absence of scatterings. We obtain%

\begin{align}
&  \frac{\partial}{\partial t}\mathcal{\vec{M}}\nonumber\\
&  =\mathcal{W}\left\{  \frac{1}{i\hbar}\left[  \bar{H},\mathcal{\vec{M}%
}\right]  +\frac{1}{\hbar}\mathcal{\vec{M}}\times\mathcal{\vec{B}}\right\}  ,
\label{ballistic1}%
\end{align}
which is the ballistic version of the form of Eq. (\ref{classBlochEq}).

In carrying out the LW transform, the following relations between vector cross
products and commutation or anti-commutation relations are useful,%
\begin{align*}
\vec{A}\times\vec{B}-\vec{B}\times\vec{A}  &  =\hat{I}_{i}\hat{\epsilon}%
_{ijk}\left\{  A_{j},B_{k}\right\}  ,\\
\vec{A}\times\vec{B}+\vec{B}\times\vec{A}  &  =\hat{I}_{i}\hat{\epsilon}%
_{ijk}\left[  A_{j},B_{k}\right]  ,
\end{align*}
where $\hat{I}_{i}$ is the unit dyadic symmetric tensor or \textit{idemfactor}%
, and $\hat{\epsilon}_{ijk}$ is the anti-symmetric unit tensor. For the
purpose of making gradient expansion to make contact with classical
expressions, we give here the following for convenience the LW transform of a
commutator $\left[  A,B\right]  $ and an anticommutator $\left\{  A,B\right\}
$ in terms of Poisson bracket differential operator, $\Lambda$, as%
\begin{align}
\left[  A,B\right]  \left(  p,q\right)   &  =\cos\Lambda\left[  a\left(
p,q\right)  b\left(  p,q\right)  -b\left(  p,q\right)  a\left(  p,q\right)
\right] \nonumber\\
&  -i\sin\Lambda\left\{  a\left(  p,q\right)  b\left(  p,q\right)  +b\left(
p,q\right)  a\left(  p,q\right)  \right\}  ,\label{rnc6.5}\\
\left\{  A,B\right\}  \left(  p,q\right)   &  =\cos\Lambda\left\{  a\left(
p,q\right)  b\left(  p,q\right)  +b\left(  p,q\right)  a\left(  p,q\right)
\right\} \nonumber\\
&  -i\sin\Lambda\left[  a\left(  p,q\right)  b\left(  p,q\right)  -b\left(
p,q\right)  a\left(  p,q\right)  \right]  , \label{rnc6.6}%
\end{align}
where $\Lambda=\frac{\hbar}{2}\left(  \frac{\partial^{\left(  a\right)  }%
}{\partial p}\cdot\frac{\partial^{\left(  b\right)  }}{\partial q}%
-\frac{\partial^{\left(  a\right)  }}{\partial q}\cdot\frac{\partial^{\left(
b\right)  }}{\partial p}\right)  $. Thus if the LW transforms are not
matrices, then to lowest order, we have%
\begin{align}
\left[  A,B\right]  \left(  p,q\right)   &  =-i\frac{\hbar}{2}\left(
\frac{\partial a\left(  p,q\right)  }{\partial p}\cdot\frac{\partial b\left(
p,q\right)  }{\partial q}-\frac{\partial a\left(  p,q\right)  }{\partial
q}\cdot\frac{\partial b\left(  p,q\right)  }{\partial p}\right)
,\label{rnc6.5lead}\\
\left\{  A,B\right\}  \left(  p,q\right)   &  =\left\{  a\left(  p,q\right)
b\left(  p,q\right)  +b\left(  p,q\right)  a\left(  p,q\right)  \right\}
=2a\left(  p,q\right)  b\left(  p,q\right)  . \label{rnc6.6lead}%
\end{align}

If we compare Eqs. (\ref{singleBand})-(\ref{equivEq2}) with the particle
quantum transport equation of Eq. (\ref{QTE}), without the
\textit{zitterbewegung }$\left[  \Sigma^{<},\operatorname{Re}G^{r}\right]  $
term, we see that the leading terms of the first line of the RHS of Eq.
(\ref{equivEq2}), corresponds to the classical BTE. The next three lines give
the torque, due to spin-orbit coupling (included in the effective magnetic
field), and terms coming from the spin-dependent relaxation and dephasing
scattering mechanisms. The last line explicitly involves coupling to the total charge.

Carrying out the $\mathcal{W}$ operation in Eq. (\ref{magTransportEq2}), we
find that the leading gradient terms in the RHS reads%
\begin{align}
&  \left(  \frac{\partial}{\partial t_{1}}+\frac{\partial}{\partial t_{2}%
}\right)  \mathcal{\vec{M}}\nonumber\\
&  =-\frac{1}{m}\left(  \vec{p}+\frac{e}{c}\vec{A}\left(  \vec{q}\right)
\right)  \cdot\nabla_{q}\mathcal{\vec{M}}-e\left(  \vec{E}+\frac{1}{2c}\left(
\vec{v}\times\vec{B}\right)  \right)  \cdot\nabla_{p}\mathcal{\vec{M}}%
-\frac{1}{\hbar}\bar{\Gamma}\mathcal{\vec{M}}+\frac{e}{mc}\bar{\Sigma}%
^{<}\mathcal{\vec{A}+}\frac{1}{\hbar}\mathcal{\vec{B}}\times\mathcal{\vec{M}%
}\nonumber\\
&  -\frac{1}{2}\left[  \frac{\partial\operatorname{Re}\bar{\Sigma}^{r}%
}{\partial p}\frac{\partial\mathcal{\vec{M}}}{\partial q}-\frac{\partial
\operatorname{Re}\bar{\Sigma}^{r}}{\partial q}\frac{\partial\mathcal{\vec{M}}%
}{\partial p}\right]  -\frac{1}{2}\left[  \frac{\partial\bar{\Sigma}^{<}%
}{\partial p}\frac{\partial\operatorname{Re}\mathcal{\vec{M}}^{r}}{\partial
q}-\frac{\partial\bar{\Sigma}^{<}}{\partial q}\frac{\partial\operatorname{Re}%
\mathcal{\vec{M}}^{r}}{\partial p}\right] \nonumber\\
&  +\frac{1}{\hbar}\operatorname{Re}\vec{\Xi}^{r}\times\mathcal{\vec{M}}%
+\frac{1}{\hbar}\vec{\Xi}^{<}\times\operatorname{Re}\mathcal{\vec{M}}%
^{r}\nonumber\\
&  -\frac{1}{2}\hat{I}_{i}\hat{\epsilon}_{ijk}\left[  \frac{\partial\gamma
_{j}}{\partial p}\frac{\partial\mathcal{M}_{k}}{\partial q}-\frac
{\partial\gamma_{j}}{\partial q}\frac{\partial\mathcal{M}_{k}}{\partial
p}\right]  +\frac{1}{2}\frac{e}{2mc}\hat{I}_{i}\hat{\epsilon}_{ijk}\left[
\frac{\partial\Xi_{j}}{\partial p}\frac{\partial\mathcal{A}_{k}}{\partial
q}-\frac{\partial\Xi_{j}}{\partial q}\frac{\partial\mathcal{A}_{k}}{\partial
p}\right] \nonumber\\
&  -\frac{1}{2}\left[  \frac{\partial\mathcal{\vec{B}}}{\partial p}%
\frac{\partial M_{o}^{<}}{\partial q}-\frac{\partial\mathcal{\vec{B}}%
}{\partial q}\frac{\partial M_{o}^{<}}{\partial p}\right]  -\frac{1}{\hbar
}\vec{\gamma}M_{o}^{<}-\frac{1}{\hbar}\vec{\Xi}^{<}\operatorname{Im}M_{o}%
^{r}\nonumber\\
&  -\frac{1}{2}\left[  \frac{\partial\operatorname{Re}\vec{\Xi}^{r}}{\partial
p}\frac{\partial M_{o}^{<}}{\partial q}-\frac{\partial\operatorname{Re}%
\vec{\Xi}^{r}}{\partial q}\frac{\partial M_{o}^{<}}{\partial p}\right]
-\frac{1}{2}\left[  \frac{\partial\vec{\Xi}^{<}}{\partial p}\frac
{\partial\operatorname{Re}M_{o}^{r}}{\partial q}-\frac{\partial\vec{\Xi}^{<}%
}{\partial q}\frac{\partial\operatorname{Re}M_{o}^{r}}{\partial p}\right]  ,
\label{smqte}%
\end{align}
where all quantities above are LW transforms, i.e., they are functions defined
in phase space $\left(  p,q,t\right)  =\left(  \vec{p},-E,\vec{q},t\right)  $.
In the first line of Eq. (\ref{smqte}) and the first term of the second line
arise from the relation,
\begin{align}
&  -\frac{1}{2}\left[  \frac{\partial\left(  \bar{H}+\operatorname{Re}%
\bar{\Sigma}^{r}\right)  }{\partial p}\frac{\partial\mathcal{\vec{M}}%
}{\partial q}-\frac{\partial\left(  \bar{H}+\operatorname{Re}\bar{\Sigma}%
^{r}\right)  }{\partial q}\frac{\partial\mathcal{\vec{M}}}{\partial p}\right]
\nonumber\\
&  =-\frac{1}{m}\left(  \vec{p}+\frac{e}{c}\vec{A}\left(  \vec{q}\right)
\right)  \cdot\nabla_{q}\mathcal{\vec{M}}-e\left(  \vec{E}+\frac{1}{2c}\left(
\vec{v}\times\vec{B}\right)  \right)  \cdot\nabla_{p}\mathcal{\vec{M}%
}\nonumber\\
&  -\frac{1}{2}\left[  \frac{\partial\operatorname{Re}\bar{\Sigma}^{r}%
}{\partial p}\frac{\partial\mathcal{\vec{M}}}{\partial q}-\frac{\partial
\operatorname{Re}\bar{\Sigma}^{r}}{\partial q}\frac{\partial\mathcal{\vec{M}}%
}{\partial p}\right]  , \label{singleHterm}%
\end{align}
where the RHS arise from the expression given by the first terms of Eq.
(\ref{hamiltonian}).We have used the gauge: $\vec{A}\left(  q\right)  =\left(
\frac{1}{2}\vec{B}\times\vec{q}\right)  $, $\vec{B}$ is the uniform external
magnetic field\cite{derivQH}.

The first line in the RHS of Eq. (\ref{smqte}) is the Boltzmann equation in a
magnetic field with effects of spin-orbit coupling. The second line is the
contribution to the Boltzman equation if the effects of $\operatorname{Re}%
\bar{\Sigma}^{r}$ and $\operatorname{Re}\mathcal{\vec{M}}^{r}$ are included.
The rest of this equation incorporates spin-dependent scattering,
$\operatorname{Re}\vec{\Xi}^{r}$, which directly contribute to the torque. The
precession of the magnetic moment, $\mathcal{\vec{M}}$, is also correlated
with the torque exerted by $\mathcal{\vec{\gamma}}$, and similarly for
'\textit{scattering-in}' precession (fluctuation) of $\mathcal{\vec{A}}$ due
to torque exerted by $\vec{\Xi}$. The last two lines describe the coupling to
the total charge. Thus, spin-dependent scattering terms are capable of
describing the \textit{spatio--temporal-dependent} scattering dynamics of
various mechanisms of spin relaxations and precession, namely, Elliott-Yafet
(spin-dependent impurities and phonon scatterings), D'yakonov-Perel' (due to
electron-electron scatterings, with electrons executing random-walk in
momentum space, resulting in the fluctuations of magnitude and direction of
spin precession axis). The Bir-Aronov-Pikus scattering (due to the interaction
with holes in valence band, serving as a spin sink), and the hyperfine
interactions are not relevant in our present treatment.

Until now the treatments of spin transport are mainly focused on scattering
mechanisms, based on quasi-classical steady-state conditions and/or where
nonuniformity in real space is often not considered, common in micromagnetics.
On the other hand for spintronic transistor devices, strong inhomogeneity in
real space is of prime considerations. The calculations of the self energy
$\Sigma$ often have the goal of obtaining the relaxation times to be used in
classical kinetic equations.

A calculation where the self-energy spinor due to many-body effects of
spin-orbit coupling is treated has been given by Rajagopal\cite{rajagopal}. A
first principle \textit{ab initio} calculation of the self-energies has been
given by Zhukov and Chulkov\cite{zhukov} for the Elliot-Yafet mechanism in Al,
Cu, Au, Nb, and Ta, and by Mower, Vignale, and Tokatly\cite{mower} for
D'yakonov-Perel spin relaxation mechanism in photo-excited electron-hole
liquid in intrinsic semiconductors exhibiting spin-split band structure.

Our derivation of the nonequilibrium spin magnetization QDF transport
equations is based on the separation of charge and spin as the relevant
variables. The resulting SMQDFT equations should provide the fundamental basis
for carrying out the numerical simulation of the switching or time-dependent
performance analyses of spintronic devices, such as those that make use of
thin insulating layers between conducting metal layers\cite{mismatch}.
Potential barriers of up to $1\ eV$ exist in promising spintronic devices,
exemplified by the spin valve and magnetic tunnel transistors\cite{zutic,
jansen}.

\medskip\medskip

One of the authors (FAB) is grateful to Prof. Bro. Romel G. Bacabac, Prof.
Raymund Sarmiento, the physics and biology faculty for their hospitality
during his visit at the Department of Physics, University of San Carlos at
Talamban, Cebu City, Philippines.

\section{}


\begin{thebibliography}{99}
\bibitem {contrast}In contrast to Fourier transformation, the matrix element
in the integrand of Eq. (\ref{preq142}) which is diagonal in either lattice
position or crystal momentum is invariant under the lattice Weyl
transformation. Clearly, the crystal momentum and lattice position themselves,
considered as matrx element of momentum and position operator, respectively,
are invariant under lattice Weyl transformation.
\bibitem {boltzmann}That the gradient expansion of Eq. (\ref{QTE}), with the
$\left[  \Sigma^{<},\operatorname{Re}G^{r}\right]  $ term neglected, yields
the classical Boltzmann transport equation for spinless system is often viewed
as justifying the neglect of $\left[  \Sigma^{<},\operatorname{Re}%
G^{r}\right]  $ in the quantum transport equations\cite{kadanoff}.
\bibitem {note1}It is worth mentioning that the transformation from four
independent operators or quantum states to their four linearly-independent
combinations has yielded some of the deepest concepts of quantum mechanics,
namely, spin angular momentum and quantum teleportation. A two-state
eigenvector space, denoted by basis $\left\vert \uparrow\right\rangle $ and
$\left\vert \downarrow\right\rangle $, has the following complete set of four
basis operators, $\hat{b}_{\downarrow\uparrow}=\left\vert \downarrow
\right\rangle \left\langle \uparrow\right\vert $, \ $\hat{b}_{\downarrow
\uparrow}^{\dagger}=\left\vert \uparrow\right\rangle \left\langle
\downarrow\right\vert $, \ $\hat{d}_{\downarrow\downarrow}^{\dagger
}=\left\vert \uparrow\right\rangle \left\langle \uparrow\right\vert $,
\ $\hat{d}_{\downarrow\downarrow}=\left\vert \downarrow\right\rangle
\left\langle \downarrow\right\vert .$The spin operators are obtained as simple
linear combinations of the above basis operators as, $\hat{S}_{o}=\frac{\hbar
}{2}\left(  \hat{d}_{\downarrow\downarrow}^{\dagger}+\hat{d}_{\downarrow
\downarrow}\right)  $, $\hat{S}_{x}=\frac{\hbar}{2}\left(  \hat{b}%
_{\downarrow\uparrow}^{\dagger}+\hat{b}_{\downarrow\uparrow}\right)  $,
$\hat{S}_{y}=\frac{\hbar}{2i}\left(  \hat{b}_{\downarrow\uparrow}^{\dagger
}-\hat{b}_{\downarrow\uparrow}\right)  $, $\hat{S}_{z}=\frac{\hbar}{2}\left(
\hat{d}_{\downarrow\downarrow}^{\dagger}-\hat{d}_{\downarrow\downarrow
}\right)  ,$ obeying the quantum-mechanical spin angular-momentum commutation
relations, namely, $\left[  \hat{S}_{j},\hat{S}_{k}\right]  =i\hbar
\epsilon_{jkl}\hat{S}_{l}$, \ \ $i,j,k\neq o$. In matrix form, these are
represented by the well-known Pauli matrix, $\vec{\sigma}$, namely, $\hat
{S}_{x}=\frac{\hbar}{2}\sigma_{x}$, $\hat{S}_{y}=\frac{\hbar}{2}\sigma_{y}$,
$\hat{S}_{z}=\frac{\hbar}{2}\sigma_{z}$. \ Note that operator $\hat{S}_{o}$
represent the identity operator (by virtue of the completeness relation for
the eigenvectors; statistically referred to as a pure state) multiplied by
$\frac{\hbar}{2}$. Any $2\times2$ matrix in 'spin' space can be expanded in
terms of the Pauli and identity operators.
\bibitem {note2}We note further that the linear combination of product states
in a two-qubit system generates the entanglement Bell basis states which play
a crucial role in the theory of quantum teleportation\cite{buot2,
nielsen-chuang}.
\bibitem {dresRashba}For example, for symmetric quantum well, the Rashba
spin-orbit coupling can be expressed as $H_{R}=\frac{\alpha}{\hbar}\left(
\hat{e}_{z}\times\vec{p}\right)  \cdot\vec{\sigma}$, where $\alpha$ is the
Rashba spin-orbit parameter.
\bibitem {derivQH}In taking the derivative of $\frac{\partial\bar{H}}{\partial
q}$, we make the expansion of kinetic energy terms in $\bar{H}\Rightarrow
\frac{\vec{p}^{2}}{2m}+\frac{e}{mc}\vec{p}\cdot\vec{A}+\left(  \frac{e}%
{mc}\right)  ^{2}\vec{A}\cdot\vec{A}$ using the symmetric Landau gauge:
$\vec{A}=\frac{1}{2}\left(  \vec{B}\times\vec{q}\right)  $. Then in taking the
gradient with respect to $\vec{q}$, the $\vec{A}\cdot\vec{A}$ term does not
contribute and we are left with,%
\begin{align*}
\frac{e}{mc}\vec{p}\cdot\frac{\partial}{\partial\vec{q}}\vec{A}  &  =\frac
{e}{mc}\vec{p}\cdot\frac{\partial}{\partial\vec{q}}\frac{1}{2}\left(  \vec
{B}\times\vec{q}\right) \\
&  =-\frac{e}{mc}\frac{1}{2}\vec{p}\cdot\frac{\partial}{\partial\vec{q}%
}\left(  \vec{q}\times\vec{B}\right)  =-\frac{e}{mc}\frac{1}{2}\vec{p}%
\cdot\left(  \hat{I}\times\vec{B}\right) \\
&  =-\frac{e}{mc}\frac{1}{2}\left(  \vec{p}\times\vec{B}\right)  =-\frac
{e}{2c}\left(  \vec{v}\times\vec{B}\right)
\end{align*}
\bibitem {mismatch}This heterostucture configuration solves the mismatch
problem between the conductivities of metal layers which has hindered
achieving high spin-injection efficiency.
\end{thebibliography}
\end{document}